      [1999/12/01 v1.4c Il Nuovo Cimento]
\documentclass{cimento}
\usepackage{epsfig}
\newcommand{\ud}{\mathrm{d}}
\begin{document}
\title{A rigorous definition of mass in special relativity}
\author{
E.~Zanchini\from{ins:x} \atque A.~Barletta\from{ins:x}}
\instlist{\inst{ins:x} Universit\`{a} di Bologna - Dipartimento di
Ingegneria Energetica, Nucleare e del Controllo Ambientale, Viale
Risorgimento 2 - 40136 Bologna, Italy}
\PACSes{\PACSit{01.55.+b}{General physics}
\PACSit{45.20.D-}{Newtonian mechanics}
\PACSit{03.30.+p}{Special relativity}\PACSit{45.50.-j}{Dynamics
and kinematics of a particle and a system of particles}}

\maketitle

\begin{abstract}
The axiomatic definition of mass in classical mechanics, outlined
by Mach in the second half of 19th century and improved by
several authors, is simplified and extended to the theory of
special relativity. According to the extended definition presented
here, the mass of a relativistic particle is independent of its
velocity and coincides with the rest mass, i.e., with the mass
defined in classical mechanics. Then, force is defined as the
product of mass and acceleration, both in the classical and in the
relativistic framework.
\end{abstract}


\section{Introduction}
The problem of stating a rigorous definition of mass in classical mechanics and special relativity has not yet received a widely shared solution. Several university textbooks still employ the traditional definition of mass based on the concept of force. Some treatments of this kind, chosen among the most accurate, will be briefly commented.

Halliday, Resnick and Walker \cite{ref:Halliday} define force as an interaction that can cause acceleration and a \textit{unit of force} as the force which causes a unit acceleration when applied to a reference body $R$, to which the value $m_R = 1$ of a property $m$ called mass is assigned. By definition, the magnitude of a force \textit{applied to R} equals the magnitude of the acceleration that it produces. In order to measure the mass $m_A$ of a body $A$, one should apply \textit{the same force} to $A$ and to $R$ and measure the ratio of the magnitudes of the accelerations of $R$ and $A$. A similar argument is presented by Beiser \cite{ref:Beiser}. Indeed, by this method forces are defined quantitatively only when applied to $R$; a general rule to establish when the same force is applied to different bodies, such as $R$ and $A$, is not stated; moreover, one should prove that the ratio of two masses, $m_B/m_A$, is independent of the choice of the reference body $R$.

The treatment presented by Wellner \cite{ref:Wellner} is as follows. One chooses a reference object with unit mass, called \textit{standard mass}; one also states that a fraction $1/q$ (in volume) of the standard mass has mass $m = 1/q$. Then one defines force as the product of mass and acceleration, namely $\textbf{f} = m \textbf{a}$. At this stage mass, and thus force, is defined only for the standard mass and parts thereof. By means of the standard mass and of the definition of force, one calibrates a spring scale. Thus, force is defined (for any body) by the position of an index in the calibrated spring scale. By employing the calibrated spring scale, one measures the magnitude $f_A$ of the force which acts on a body $A$ and hence the mass $m_A$ of the body as $m_A = f_A/a_A$, where $a_A$ is the magnitude of the acceleration of $A$. Indeed, one should prove: that the procedure is reproducible; that the mass $m_A$ of a body $A$ is independent of both the position and the velocity of $A$; that the ratio of two masses, $m_B/m_A$, is independent of the choices of both the spring scale and the standard mass.\\
\indent As clearly pointed out by Jammer \cite{ref:jammer} and by Lindsay and Margenau \cite{ref:Lindsay-Margenau}, any definition of mass which employs the concept of force is vitiated by a logical circularity, because any rigorous and general definition of force is based on the concept of mass. As a consequence, the definition of mass must be founded only on kinematic quantities. Similarly, the statement of the inertia
principle and the definition of an inertial reference frame cannot
involve the concept of force. \\
\indent A definition of mass based only on kinematic quantities
was conceived by Mach \cite{ref:mach}, in the last decades of 19th
century, and has been improved by several authors
\cite{ref:Lindsay-Margenau},
\cite{ref:Bressan}--\cite{ref:guerra}. Thus, a rigorous definition
of mass in classical mechanics, which will be
called \textit{axiomatic definition of mass}, is now available; in some textbooks, it is outlined after a preliminary traditional treatment \cite{ref:Alonso}, \cite{ref:Ohanian}. \\
\indent The axiomatic definition of mass is stated along the
following lines. An inertial reference frame is defined, without
employing the concept of force, and the existence of inertial
reference frames is stated (first law of dynamics). Then, the
second law of dynamics is stated as follows
\cite{ref:Lindsay-Margenau}, \cite{ref:Benvenuti}.\\
\indent Let $A$ and $B$ be two material particles placed far away
from all the others, so that they can be considered as isolated;
let $\textbf{a}_{AB}$ be the acceleration of $A$ due to $B$, with
magnitude $a_{AB}$, and let $\textbf{a}_{BA}$ be the acceleration
of $B$ due to $A$, with magnitude $a_{BA}$. Then,
$\textbf{a}_{AB}$ and $\textbf{a}_{BA}$ are opposite vectors and
the ratio $m_{BA}$ = $a_{AB}$/$a_{BA}$ is a constant, which is
called mass ratio of $B$ with respect to $A$. Moreover, if three
pairs of isolated material particles, $(A,B)$, $(A,C)$ and $(B,C)$
are considered, the mass ratios $m_{BA}$, $m_{CA}$ and $m_{AB}$
fulfil the equation
\begin{equation}\label{mratios}
m_{BC} = m_{BA} \, m_{AC}= \frac{m_{BA}}{m_{CA}}\;\;.
\end{equation}
On account of the second law, the mass $m_B$ of $B$ can be defined
as follows. Let us consider $A$ as a reference material particle,
to which an arbitrarily chosen value of mass, $m_A$, is assigned.
Then, the mass $m_B$ of $B$ is defined as
\begin{equation}\label{Machmass}
m_B = m_A\,m_{BA} \;\;.
\end{equation}
As a consequence of Eqs. (\ref{mratios}) and (\ref{Machmass}), the
ratio $m_B/m_C$ between the mass of $B$ and that of $C$ is
independent of the choice of $A$ and can be measured directly,
i.e.
\begin{equation}\label{Machmass3}
m_B/m_C = m_{BC}\;\;.
\end{equation}

\indent The definition outlined above is rigorous. However, it is
neither universally considered as the true definition of mass nor
widely adopted in textbooks. In our opinion, the axiomatic
definition of mass has obtained only a
partial success for the following reasons.\\
\indent a) The axiomatic statement of the second law of dynamics
is rather involved and looks more mathematical than physical;
moreover,
the whole formalism appears as difficult for didactic presentations.\\
\indent b) The axiomatic statement of the second law of dynamics
does not hold in the framework of special relativity. In fact, the
Lorentz electromagnetic force does not fulfil the third law of
dynamics \cite{ref:assis}, \cite{ref:Feynman}. Although some
authors still support an alternative theory of electrodynamics
which fulfils the third law \cite{ref:Pappas}, it is widely
accepted that the standard theory of electrodynamics is correct,
as is confirmed also by a recent experiment \cite{ref:Cavalleri1}.
Thus, for a pair ($A,\,B$) of isolated and electrically charged
particles, if the magnitude of the relative velocity of the
particles is comparable with light speed, the accelerations
$\textbf{a}_A$ and $\textbf{a}_B$ of the particles have not, in
general, opposite directions. As a consequence, the axiomatic
statement of the second law of dynamics does not hold for a pair
of relativistic material points.\\
\indent In the present paper, the axiomatic definition of mass is
simplified and extended to the special theory of relativity. In
the framework of classical mechanics, the definitions of inertial
reference frame and of isolated pair of material particles are
improved and the definition of mass is simplified. Then, the mass
of a material point $A$ endowed with a relativistic speed $v$ with
respect to an inertial reference frame $O$ is defined by
considering a pair of isolated material points $(A,\,B)$ such
that, at an instant $t$, $B$ has the same classical velocity
$\textbf{v}$ as $A$. It is proved that, at time $t$, the
four-acceleration of $A$ is proportional to that of $B$, namely
$\alpha_{A\nu} = - k \,\alpha_{B\nu}$, and that the positive
constant $k$ is independent of $\textbf{v}$, $i.e.$, has the same
value same as in the classical limit. The mass of $A$ is defined
as $m_A = m_B/k$, where $m_B$ is the classical mass of $B$.
According to this definition, the mass of $A$ in special
relativity coincides with the classical mass of $A$, in agreement
with recent investigations on the concept of
mass in relativistic physics \cite{ref:sachs2}--\cite{ref:Okun2}.\\
\indent Both in the classical and in the relativistic framework,
force is defined as the product of
mass and acceleration.\\

\section{Mass and force in non-relativistic classical mechanics}

\subsection{Definition: inertial reference frame}
Let $A$ be an arbitrarily chosen material point, whose motion is
observed with respect to a reference frame $O$. If the velocity of
$A$ is constant whenever $A$ is placed far away from any other
physical object, then $O$ will be called an inertial reference
frame.

\subsection{First law of dynamics}
Inertial reference frames exist.

\subsection{Conditions and symbols}
In the whole parer, only inertial reference frames will be considered. In this section,
we will assume that the speed $v$ of every
material point is much smaller than light speed $c$. The
Galilean transformation of coordinates ensures that, if $O$ is an
inertial reference frame and $O'$ is any reference frame which
moves with a constant velocity with respect to $O$, then $O'$ is
an inertial reference frame as well. We will denote by
$\textbf{a}_A$ the acceleration of a material point $A$ and by
$a_A$ the magnitude of $\textbf{a}_A$.

\subsection{Definition: isolated
system of material points} Let $(A,\,B,\,C,\ ... )$ be a set of
material points. If each material point of the set has a constant
velocity whenever all the others are removed and placed far away
from the region of space considered, then $(A,\,B,\,C,\ ... )$
will be called an isolated system of material points.

\subsection{Second law of dynamics}
Let $(A,\,B)$ be an isolated pair of material points. At any time
instant $t$, if $\textbf{a}_A$ is non--vanishing also
$\textbf{a}_B$ is non--vanishing; moreover, $\textbf{a}_A$ and
$\textbf{a}_B$ are parallel to the straight line from $A$ to $B$,
with opposite directions, and the ratio $a_A/a_B$ is a constant
determined uniquely by the choice of $A$ and $B$ ($i.e.$,
independent of the positions and the velocities of $A$ and $B$).

\subsection{Definition: mass of a material point}
Let us consider an isolated pair of material points $(A,\,R)$,
where $R$ is a reference material point, and a time instant $t$
such that $\textbf{a}_R$ is non-vanishing. We will call mass of
$A$ the quantity $m_A$ defined as follows:
\begin{equation}\label{mass}
m_A = m_R \,\frac{a_R}{a_A}\;\;,
\end{equation}
where $m_R$ is a positive real number that will be called mass of
$R$. Since $R$ and $m_R$ are fixed once and for all, the second
law of dynamics ensures that $m_A$ has a unique value, which is
strictly positive.

\subsection{Comment}
Clearly, at this stage the definition of mass is incomplete. The
mass of a material point could be measured only by employing the
reference material point $R$.

\subsection{Third law of dynamics}
Let $(A,\,B)$ be an isolated pair of material points and let $m_A$
and $m_B$ be the masses of $A$ and of $B$, measured with respect
to a reference material point $R$. Then, at any time instant,
\begin{equation}\label{resultant}
m_A\, \textbf{a}_A + m_B\,\textbf{a}_B = 0\;\:.
\end{equation}

\subsection{Direct measurement of the ratio of two masses}
Let $(A,\,B)$ be an isolated pair of material points. Let $m_A$
and $m_B$ be the masses of $A$ and of $B$. Then, at any time
instant $t$ chosen so that $a_B$ is non-vanishing, Eq.
(\ref{resultant}) implies that
\begin{equation}\label{massratios}
\frac{m_A}{m_B} = \frac{a_B}{a_A}\;\;.
\end{equation}

\subsection{Comment}
We have proved that the ratio $m_A/m_B$ between the mass of $A$
and that of $B$ is independent of the choice of the reference
material point and can be measured directly. Thus, the definition
of mass of a material point has been completed.

\subsection{Definition: force which acts on a material point}
Let $A$ be a material point with mass $m_A$. We will call force
which acts on $A$, at a time instant $t$, the vector
\begin{equation}\label{force}
\textbf{f}_A = m_A \,\textbf{a}_A\;\;,
\end{equation}
where $\textbf{a}_A$ is the acceleration of $A$ with respect to an
inertial reference frame $O$ at the instant $t$.

\section{Mass and force in special relativity}

Let us denote the time coordinate by $x_0 = c t$, where $t$ is
time and $c$ is light speed in free space, and the space
coordinates by $x_i \; (i = 1, 2, 3)$. Let $A$ be a material point
in motion with respect to a reference frame $O$. At any time
instant $t$, one can measure the speed $v$ of $A$ with respect to
$O$,
\begin{equation}\label{speed}
v = \left[\left( \frac{\ud x_1}{\ud t}\right)^2 +
\left(\frac{\ud x_2}{\ud t}\right)^2
+ \left(\frac{\ud x_3}{\ud t}\right)^2\right]^{1/2} \;\;,
\end{equation}
as well as the dimensionless parameters
\begin{equation}\label{parameters}
\beta = \frac{v}{c}\;\;,\;\; \gamma = \left(1 - \beta^2\right)^{-1/2} \;\;.
\end{equation}
A time interval $\ud t$, measured by an observer at rest with
respect to $O$, corresponds to a proper-time interval $\ud \tau$,
measured by an observer at rest with respect to $A$, given by
\begin{equation}\label{propertime}
d \tau =  \frac{dt}{\gamma}  \;\;.
\end{equation}
We will call velocity of $A$, at an instant $t$, the four-vector
\begin{equation}\label{velocity}
u_\mu =  \frac{\ud x_\mu}{\ud \tau} \;\;,
\end{equation}
whose components are given by
\begin{equation}\label{velcomp}
 u_0 = \gamma c \;\; , \;\; u_i =  \gamma \frac{\ud x_i}{\ud t} \;\;.
\end{equation}
We will call acceleration of $A$, at an instant $t$, the four-vector
\begin{equation}\label{acceleration}
\alpha_\mu =  \frac{\ud u_\mu}{\ud \tau} \;\;,
\end{equation}
whose components are given by
\begin{equation}\label{acccomp}
 \alpha_0 = \frac{\mathbf{v}\cdot\mathbf{a}}{c}\;\gamma^4\;\;,
\;\; \alpha_i = \gamma^2 \left(a_i +
\frac{\mathbf{v}\cdot\mathbf{a}}{c^2}\; \gamma^2 v_i \right)\;\;,
\end{equation}
as is easily obtained from standard expressions \cite{ref:rindler}
through the equality
\begin{equation}\label{dergamma}
\frac{\ud \gamma}{\ud t} = \frac{\mathbf{v}\cdot\mathbf{a}}{c^2}
\gamma^3 \;\; .
\end{equation}
In Eq. (\ref{acccomp}), $\mathbf{v}$ is the three--dimensional
classical velocity, with components $v_i = \ud x_i/\ud t$, and
$\mathbf{a}$ is the
three--dimensional classical acceleration, with components $a_i = \ud^2 x_i/\ud t^2$.\\
\indent If $O'$ is a reference frame which moves with a constant
classical velocity $\textbf{\textbf{v}}$ with respect to a
reference frame $O$, the space-time coordinates $x'_{\mu}$ of
events observed by $O'$ can be obtained from the coordinates $x_{\mu}$ of
the same events observed by $O$ by means of the Lorentz transformation
\begin{equation}\label{Lorentz}
x'_{\mu} = L_{\mu \nu} \; x_{\nu} \;\; ,
\end{equation}
where the transformation coefficients $L_{\mu \nu}$ are constants.
By differentiating twice Eq. (\ref{Lorentz}) with respect to
proper time, one obtains
\begin{equation}\label{Lorentzacc}
\alpha'_{\mu} = L_{\mu \nu} \; \alpha_{\nu} \;\; .
\end{equation}
\indent The definition of inertial reference frame, the first law
of dynamics and the definition of an isolated pair of material
particles can be taken from Section 2 without changes.\\
\indent Equation (\ref{Lorentzacc}) ensures that, if $O$ is an
inertial reference frame and $O'$ is any reference frame which
moves with a constant velocity with respect to $O$, then $O'$ is
an inertial reference frame as well. Moreover, it yields the
following corollary.

\subsection{Corollary 1}
If the four-accelerations $\alpha_{A_\mu}$ and $\alpha_{B_\mu}$ of
two material points $A$ and $B$ are proportional with respect to
an inertial reference frame $O$, $i.e. \; \alpha_{A_\mu} = k \;
\alpha_{B_\mu}$, they are proportional also with respect to a
reference frame $O'$ which moves with respect to $O$ with a
constant classical velocity $\textbf{v}$, and the proportionality
constant $k$ is the same, $i.e. \; \alpha'_{A_\mu} = k \;
\alpha'_{B_\mu}$.\\

\indent Be means of corollary 1, we will prove the validity of the
following extended statement of the second law of dynamics.

\subsection{Second law of dynamics}
Let $(A,\,B)$ be an isolated pair of material points which, at a
time instant $t$, have the same classical velocity $\textbf{v}$
with respect to an inertial reference frame $O$. Then, at time
$t$, the accelerations $\alpha_{A\nu}$ and $\alpha_{B\nu}$ are
related by the equation
\begin{equation}\label{secondlaw}
\alpha_{A\nu} = - k \, \alpha_{B\nu} \;\;,
\end{equation}
where $k$ is a positive constant determined uniquely by the choice
of $A$ and $B$ ($i.e.$, independent of the positions of $A$ and
$B$ and of their velocity $\textbf{v}$).

\subsection{Proof}
Let $A$ and $B$ be material points which, at time $t$, have the
same classical velocity $\textbf{v}$ with respect to an inertial
reference frame $O$. Let $O'$ be an inertial reference frame which
moves with respect to $O$ with the same classical velocity
$\textbf{v}$ as $A$ and $B$, $i.e.$, such that $A$ and $B$ are at
rest with respect to $O'$ at the time instant $t'$ which corresponds to $t$.\\
\indent On account of Eq. (\ref{acccomp}), the time components of
the four-accelerations of $A$ and $B$ with respect to $O'$, at the
instant $t'$, are vanishing, while the space components coincide
with the components of the classical accelerations, i.e.
\begin{equation}\label{accAOprime}
\alpha'_{A0} = 0 \;\; , \;\; \alpha'_{Ai} = a'_{Ai} \;\; .
\end{equation}
\begin{equation}\label{accBOprime}
\alpha'_{B0} = 0 \;\; , \;\; \alpha'_{Bi} = a'_{Bi} \;\; .
\end{equation}
Moreover, the classical statement of the second law of dynamics
holds with respect to $O'$ and yields
\begin{equation}\label{classicalratio}
a'_{Ai} = - k \, a'_{Bi} \; \; .
\end{equation}
Equations (\ref{accAOprime}), (\ref{accBOprime}) and
(\ref{classicalratio}) yield
\begin{equation}\label{secondlawprime}
 \alpha'_{A\mu} =  - k \; \alpha'_{B\mu} \;\; .
\end{equation}
Equation (\ref{secondlawprime}) and corollary 1 yield Eq.
(\ref{secondlaw}).

\subsection{Definition: mass of a material point}
Let $A$ be a material point with classical velocity $\textbf{v}$
with respect to an inertial reference frame $O$. Let us couple $A$
with a material point $B$ which, at time $t$, has the same
classical velocity $\textbf{v}$ as $A$ and such that the pair
$(A,\,B)$ is isolated. On account of the second law of dynamics,
at time $t$, $\alpha_{A\nu} = - k \, \alpha_{B\nu}$. We will call
mass of $A$ the quantity $m_A$ defined as follows:
\begin{equation}\label{relmass}
m_A = \frac{1}{k} \; m_B \;\;,
\end{equation}
where $m_B$ is the mass of $B$ defined in non-relativistic
classical mechanics.

\subsection{Comment}
Since $k$ is independent of velocity, $m_A$ coincides with the
mass of $A$ defined in non-relativistic classical mechanics.

\subsection{Definition: force which acts on a material point}
Let $A$ be a material point with mass $m_A$, whose motion is
observed with respect to an inertial reference frame $O$. We will
call force which acts on $A$, at a time instant $t$, the
four--vector
\begin{equation}\label{forcerel}
K_{A \nu} = m_A \,\alpha_{A \nu}\;\;,
\end{equation}
where $\alpha_{A \nu}$ is the acceleration of $A$ with respect to
$O$ at the instant $t$.

\section{Conclusions}
An axiomatic definition of mass in classical mechanics has been available for several decades; however, it is neither widely employed in textbooks nor stated in a form which applies to the broader framework of relativistic dynamics.\\
\indent In this paper, the axiomatic definition of mass has been simplified and extended to the special theory of relativity. According to the definition
proposed here, the mass of a material point $A$ in special relativity coincides with the classical mass of $A$. Both in the classical and in the relativistic framework, force has been defined as the product of mass and acceleration.\\
\indent Thus, a natural extension to special relativity of the axiomatic definitions of mass and force has been obtained.

\end{document}